# Atom Nanooptics Based on Photon Dots and Photon Holes

V. I. Balykin* and V. S. Letokhov
*Institute of Spectroscopy, Russian Academy of Sciences, Troitsk, Moscow region, 142190 Russia*
V. V. Klimov
*Lebedev Physical Institute, Russian Academy of Sciences, Leninski pr. 53, Moscow, 119991 Russia*
* balykin@isan.troitsk.ru

•

*Abstract*
New types of light fields localized in nanometer-sized regions of space were suggested and analyzed. The possibility of using these nanolocalized fields in atom optics for atom focusing and localization is discussed.

Atom optics is a new type of optics (along with photon, electron, and neutron optics). It has emerged in the last 20 years as a result of studying the action of electromagnetic radiation on spatial atomic motion [1–5]. Atom optics is subdivided into two types: (a) atom optics based on the mechanical micro- and nanostructures (e.g., zone plates) and (b) atom optics based on the use of laser radiation in the presence of static electric and magnetic fields. Atom optics encompasses physical problems that are associated with studying atomic interactions with surfaces and electromagnetic fields in an effort to determine those interaction potentials which provide a controlled action on the spatial atomic motion; among these problems are diffraction of atomic waves and their interference, focusing of atomic waves, mirror reflection, atomic localization in a confined spatial region (atomic traps), and the increase in the phase density of atomic ensembles. Laser atom optics has a number of limitations of both fundamental and technical character; since the laser fields are spatially nonlocalized, the elements of atom optics are also nonlocalized. As a result, such imperfections as aberration of atom lenses, low diffraction efficiency of atomic waves, and limiting contrast of interference fringes in atomic interferometers are inherent in the elements of atom optics.

From general physical considerations, it is clear that spatially localized potentials are more favorable for constructing the elements of atom optics. At present, only two types of spatially localized laser fields are known: (a) a surface (evanescent) light wave arising upon total internal reflection (one-dimensional light localization) and (b) a light field appearing upon its diffraction by structures with characteristic sizes smaller than the light wavelength. The most familiar example of the second type of localized light field is provided by a field arising upon the diffraction by an aperture with a size smaller than the wavelength in an ideally conducting screen. In this case, a local three-dimensional field maximum forms near a small aperture, with sizes mainly determined by the size of aperture [6–8].

A serious disadvantage of the field localized near an isolated aperture is that it is inevitably connected with the field of the associated standing wave. When an atom moves in this region, it may undergo spontaneous decay, which is highly undesirable in many problems of atom optics.

In this work, we propose new types of a spatially localized nanometer-sized light field free of the above-mentioned disadvantage. The possibility of using such a *nanofield* in the problems of atom optics ( *atom nanooptics*) is considered.

The scheme of producing spatially localized light nanofield is illustrated in Fig. 1. Two flat conducting screens with a distance between them on the order of or shorter than the light wavelength, $d \leq \lambda$, form a planar two-dimensional waveguide for laser radiation that enters the waveguide from the side. As known [9], the solutions to the Maxwell equations for a waveguide consisting of two parallel ideally conducting planes allow the propagation of radiation in a waveguide of an arbitrarily small thickness $d$, including the thickness much smaller than the radiation wavelength. The solution inside the waveguide coincides with a plane wave whose electric field is directed perpendicular to the planes. In fact, this system is a double-wire line and provides two-dimensional nanometer-sized light localization [9].

Now let two small coaxial apertures with radius $a < \lambda$ be formed in a conducting screen (Fig. 1). If the aperture diameters are much smaller than the wavelength of incident radiation, it will practically not pass through these apertures but will be strongly modified near each of the apertures. In fact, the field is reduced near the apertures in the region with the characteristic sizes on the order of the aperture diameter, i.e., much smaller than the radiation wavelength $\lambda$. The volume of this region is $V \sim a^2 d \ll \lambda^3$. A field modification of this kind can naturally be called a "*photon hole.*"

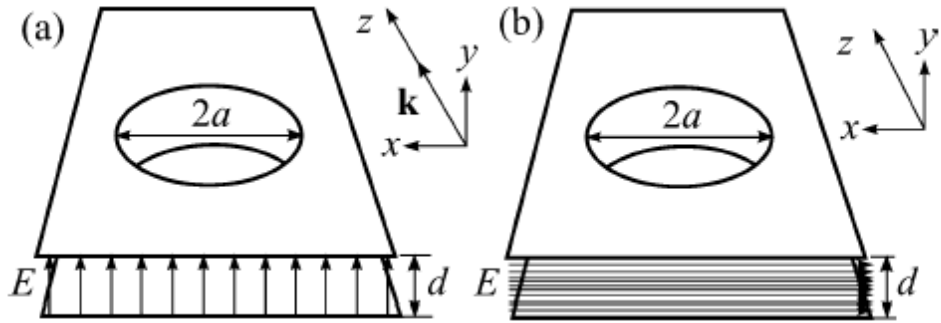

**Fig. 1.** Geometry of the formation of (a) photon hole and (b) photon dot by nanoapertures in a planar optical waveguide formed from two conducting planes.

The determination of the field distribution near the apertures in the waveguide walls is a challenging task of electrodynamics. In the particular case of a nanoaperture ($a \ll \lambda$), the problem becomes quasistatic. Indeed, let the wave propagate along the Z axis, as shown in Fig. 1a, and be polarized along the Y axis. Then, in the absence of apertures, the nonzero field components (for the traveling wave) are

$$E_y = E_0 e^{i(kz-\omega t)}$$
$$H_x = -E_0 e^{i(kz-\omega t)} \tag{1}$$

where $k = \omega / c$. In the presence of small apertures, this problem reduces, to a first approximation, to the quasistatic problem. The corresponding general solution is obtained by solving the integral equation for the charge densities in planes [10]. If the waveguide thickness $d$ is larger than the aperture diameter, $d > a$, one can ignore the mutual influence of apertures, so that the problem reduces to the superposition of fields caused by the diffraction by individual apertures. The problem of the modification of a uniform field in the presence of a conducting plane with a round aperture in it can be solved analytically [11].

The resulting expression for the potential of electric field $\mathbf{E} = -\nabla \varphi$ has the form

$$\varphi(r,y) = \varphi^+\left(r, y - \frac{d}{2}\right) + \varphi^-\left(r, y + \frac{d}{2}\right),$$

$$\varphi^{\pm}(r,z) = -\frac{E_0}{2}(z \mp d) \pm \frac{E_0}{\pi}|z|\left(\frac{1}{\mu(r,z)} + \text{arctg}\,\mu(r,z)\right), \tag{2}$$

$$\mu(r,z) = \sqrt{\frac{1}{2}\left(\frac{r^2}{a^2} + \frac{z^2}{a^2} - 1\right) + \frac{1}{2}\sqrt{\left(\frac{r^2}{a^2} + \frac{z^2}{a^2} - 1\right)^2 + 4\frac{z^2}{a^2}}}$$

where $r^2 = x^2 + z^2$.

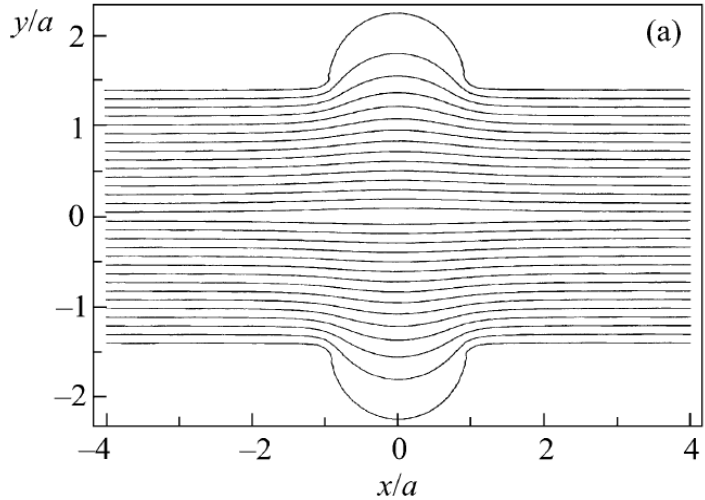

**Fig. 2.** (a) Isolines of potential (2) describing a photon hole with $a/d = 1/3$.

The distribution of the equipotential lines (2) and the electric-field energy density proportional to $\mathbf{E}^2 = (\nabla \varphi)^2$ are shown in Fig. 2. One can see in Fig. 2 that a photon hole or, to be more precise, photon saddle with the characteristic sizes determined by the aperture size and the waveguide thickness actually appears near the apertures.

We now consider one more way of field localization in nanometer-sized regions (Fig. 1b). It generalizes the method of localization near an aperture [8] but is free from the drawback associated with the presence of a standing wave. Let us again take two ideally conducting planes with apertures in them, but let the planes be separated by $d=\lambda/2$. In this case, one of the solutions ($TE_{01}$) in the absence of apertures has the nonzero components

$$E_{0,x} = -2i\cos\left(\frac{\pi}{d}y\right)e^{-i\omega t}$$
$$H_{0,z} = 2\sin\left(\frac{\pi}{d}y\right)e^{-i\omega t}$$
(3)

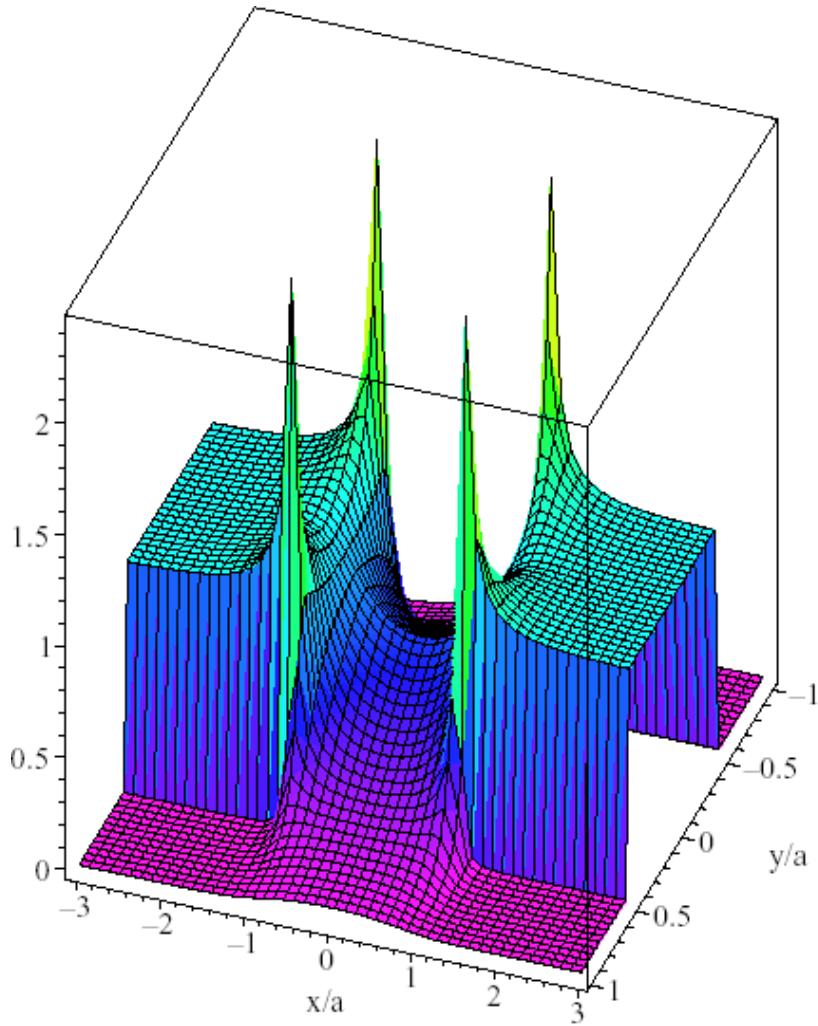

**Fig. 2.** (b) Electromagnetic-field intensity for a photon hole with $a/d = 1$.

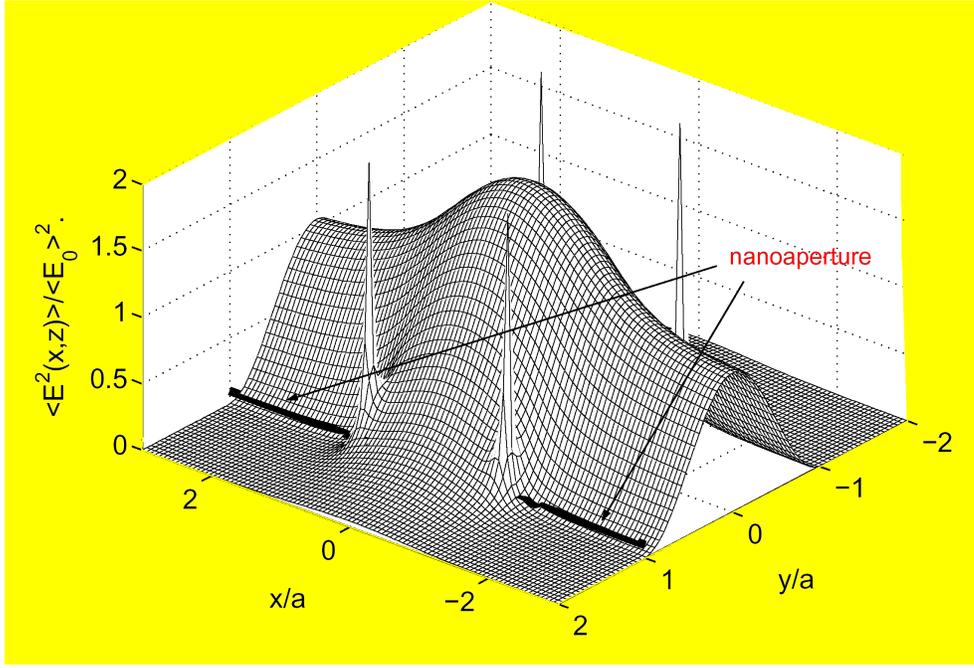

**Fig. 3.** Electromagnetic-field intensity for a photon dot with $a/d = 0.5$.

Physically, solution (3) corresponds to the standing wave whose wave vector is perpendicular (directed along the **Y** axis) to the planes. At the same time, Eq. (3) is a portion of the standing wave formed upon the reflection of a plane wave incident normally onto one of the planes. Due to the condition $d = \lambda/2$, the other plane occurs in the node of this standing wave and does not affect it.

If the apertures are small compared to the wavelength, $a \ll d = \lambda/2$, their mutual influence can be ignored, and one can use the solution to the problem of diffraction by a single aperture [6–8]. For the circular polarization, the field components are

$$\mathbf{E}_0 = -\frac{1}{\sqrt{2}}(1,0,i)\,2i\cos\frac{\pi}{d}y, \quad (-\frac{d}{2} < y < \frac{d}{2})$$
$$\mathbf{H}_0 = -\frac{1}{\sqrt{2}}(i,0,-1)\,2\sin\frac{\pi}{d}y, \quad (-\frac{d}{2} < y < \frac{d}{2}) \tag{4}$$

and the time-averaged squared electric field modified by the apertures takes the form

$$\langle E^2 \rangle = \left(\frac{ka}{3\pi}\right)^2 W \tag{5}$$

$$W = \left\{ A\left(r, \left|y + \frac{d}{2}\right|\right)^2 + (A\left(r, \left|y + \frac{d}{2}\right|\right) + B\left(r, \left|y + \frac{d}{2}\right|\right))^2 + C\left(r, \left|y + \frac{d}{2}\right|\right)^2 \right\} (y < -d/2)$$

$$W = \left\{ 18\pi^2 \frac{\cos^2\frac{\pi}{d}y}{(ka)^2} + 6\pi \frac{\cos\frac{\pi}{d}y}{ka}(2\hat{A}+\hat{B}) + \hat{A}^2 + (\hat{A}+\hat{B})^2 + \hat{C}^2 \right\} (-\frac{d}{2} < y < \frac{d}{2})$$

$$W = \left\{ A\left(r, \left|y - \frac{d}{2}\right|\right)^2 + (\hat{A}\left(r, \left|y - \frac{d}{2}\right|\right) + \hat{B}\left(r, \left|y - \frac{d}{2}\right|\right))^2 + \hat{C}\left(r, \left|y - \frac{d}{2}\right|\right)^2 \right\} \quad (y > d/2)$$

where

$$\hat{A} = A\left(r, \left|y - \frac{d}{2}\right|\right) + A\left(r, \left|y + \frac{d}{2}\right|\right)$$

$$\hat{B} = B\left(r, \left|y - \frac{d}{2}\right|\right) + B\left(r, \left|y + \frac{d}{2}\right|\right) \quad (6)$$

$$\hat{C} = C\left(r, \left|y - \frac{d}{2}\right|\right) + C\left(r, \left|y + \frac{d}{2}\right|\right)$$

and

$$A(r,y) = R^{-}\left(\frac{2a^2}{R^*} + 2 - \frac{y^2}{r^2}\right) + ya\left(\frac{R^+}{r^2} - \frac{3}{a^2}\operatorname{arctg}\left(\frac{1}{R^+}\right)\right)$$

$$B(r,y) = R^{-}\left(\frac{2y^2}{r^2} - \frac{2r^2 - y^2}{R^*}\right) + yR^{+}a\left(\frac{1}{R^*} - \frac{2}{r^2}\right) + \frac{3yr^2 R^+}{aR^*(1+R^{+2})}$$

$$C(r,y) = \frac{2arR^+}{R^*(1+R^{+2})}$$

$$R^* = \left(\left(r^2 + y^2 - a^2\right)^2 + 4a^2 y^2\right)^{1/2} ; R^{\pm} = \left(\frac{R^* \pm \left(r^2 + y^2 - a^2\right)}{2a^2}\right)^{1/2} \quad (7)$$

The intensity distribution of the field near the apertures of a planar waveguide and inside the waveguide is shown in Fig. 3 for a waveguide with a half-wavelength thickness and a aperture radius $a = \lambda/4$. It is seen from the figure that the field outside the waveguide drops rather rapidly in the direction perpendicular to the waveguide plane and has a maximum inside the waveguide; i.e., a "photon dot" arises. The characteristic size of such a photon dot is $V \sim (\lambda/2) a^2 \ll \lambda^3$. The sharp field-intensity spikes near the aperture edge appear because the conductivity of the waveguide walls is assumed to be infinite. In the waveguides with a finite wall conductivity, the spike amplitudes will be less pronounced. Of prime importance is that the maximum (measured from the level in the absence of apertures) at $x = y = 0$ is twice its value for the case of a single aperture. This is caused *by the constructive interference of the fields scattered by apertures*, allowing the use of fields lower than in the case of a single aperture.

Now we consider two examples of application of the spatially localized light field (photon dots and holes) in atom focusing and localization.

**Atom lens.** We first consider the possibility of using localized fields for focusing atomic beams by the gradient force proportional to the electric-field strength. For the positive detuning between the laser frequency and the atomic-emission frequency, an atom is expelled to weaker fields, whereas, for the negative frequency detuning, it is drawn into the region with stronger fields.

In the case of a photon hole, the nanometer-sized weak-field region is surrounded inside the waveguide by a strong field and, in the case of positive frequency detuning, an atom passing through the aperture will be drawn to the axis of the system; i.e., focusing will occur. As was mentioned above, it is quite important that the focused beams mainly move in the weak-field region, where the defocusing spontaneous-decay processes are highly improbable.

In the case a photon dot and the negative frequency detuning, atoms are drawn also to the axis of the system;, and, hence, focusing again takes place. In a photon dot, an atom moves through the strong-field region, and, thus, the spontaneous decay probability is higher than in the photon hole. However, the passage time through the nanometer region is short, so that the influence of the spontaneous decay on the focusing can again be ignored.

The theory of focusing atomic beams in the regions both with maxima and minima of an electric field is well elaborated. In particular, it was shown in [12, 13] that, in the absence of spontaneous decay, an atomic beam can be focused on an area with a diameter on the order of atomic de Broglie wavelength, which is equal to several Angströms for thermal beams. These results fully apply to our photon dots and apertures, because the spontaneous decay can be ignored in our case.

**Atomic trap.** For a system with the symmetry considered above, the optical-field configurations have extreme points where the gradient is zero. The configurations of these fields can naturally be considered as the possible trap configurations. The configuration of a photon dot is stable and is genuinely three-dimensional (Fig. 3), with the characteristic volume on the order of $a^2\lambda/2 \ll \lambda^3$. The depth of such a trap is twice the depth of a single-aperture trap [8].

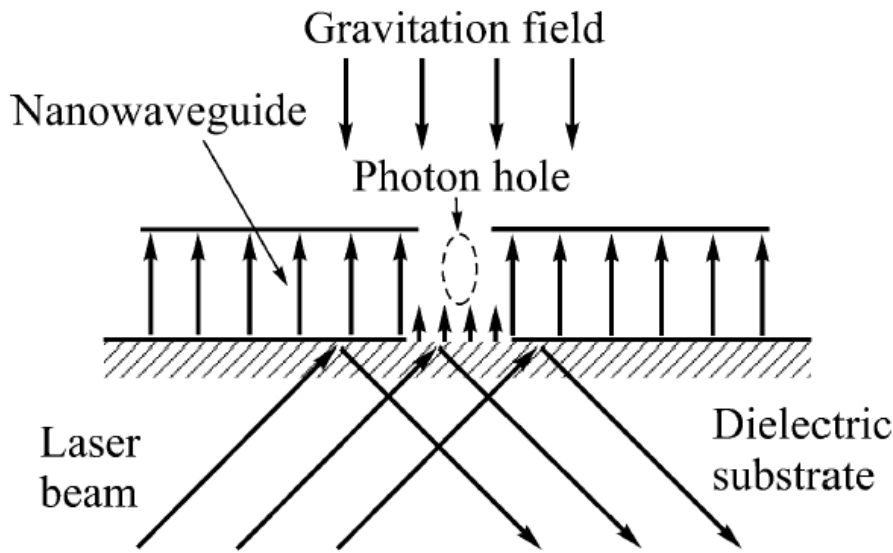

Fig.4. Atom trap configuration with a photon hole and with the use of a surface wave formed by the total internal reflection from a dielectric layer beneath the waveguide.

The extreme point of a photon hole is a saddle point (Fig. 2b): for the positive detuning, the radial motion is stable, while the motion along the axis is unstable. The opposite situation occurs for the negative detuning. However, it is desirable to determine the stable trap configuration for a nanometer-sized photon hole (all its dimensions can be much smaller than the wavelength!). The stable three-dimensional trap configuration in the case of a photon hole can be formed after slight waveguide modification with the use of the gravitational field. One such variant of a three-dimensional trap based on a photon hole is shown in Fig. 4.

In the scheme proposed, a photon hole with positive frequency detuning localizes atoms in the radial direction. From below, atoms are localized due to the exponentially decaying field that is formed upon the total internal reflection of a plane wave with the positive frequency detuning. The gravitational field provides atomic localization from above. Therefore, both optical-field configurations proposed in this work provide three-dimensional atomic localization in nanometer-sized regions. Note that one can produce a large number (lattice) of aperture pairs and, accordingly, the same number of localized fields (zero-dimensional photon holes and dots). Such a lattice allows simultaneous control of many atomic beams. In turn, such lattices can be used to form periodic lattices of localized atoms (atom lattices [14]) with a period independent of the light wavelength. The properties of such periodic lattices can be similar to the properties of planar photonic crystals [15], but, as distinct from the latter, they can combine both photon-dot lattices and lattices of localized atoms. On the whole, the approach suggested in this work forms, together with [8, 12, 13], the concept of atom nanooptics, i.e., atom optics based on optical nanofields.

We are grateful to Prof. V.G. Minogin, P.N. Melent'ev, and Prof. Shimizu (Japan) for helpful discussions of results. This work was supported in part by the Russian Foundation for Basic Research (project nos. 02-17014, 02-16337a, 02-16592a), the program "Integratsiya," and INTAS (no. INFO-00-479).